\documentclass{PoS}

\title{Peculiarities in the orbital and
precessional variability of SS433 from INTEGRAL
observations}

\ShortTitle{Hard eclipse of SS433}

\author{\speaker{A. Cherepashchuk}$^{a}$, R. Sunyaev$^b$, S. Molkov$^{b}$, E. Antokhina$^{a}$,
K. Postnov$^{a}$, A. Bogomazov$^{a}$\\
        \llap{$^{a}$}Moscow M.V. Lomonosov State University, 
Sternberg Astronomical Institute, 119992 Moscow, Russia\\
	\llap{$^{b}$}Space Research Institute of Russian Academy of Sciences, Moscow, Russia \\
        E-mail: \email{cherepashchuk@gmail.com}, \email{elant@sai.msu.ru},
\email{kpostnov@gmail.com},  \email{molkov@hea.iki.rssi.ru}}

\abstract{Based on multiyear INTEGRAL observations of SS433, a composite IBIS/ISGRI 18-60 keV orbital 
light curve is constructed around zero precessional phase $\psi_{pr}= 0$, which corresponds to a
maximum separation of the moving emission lines originated in sub-relativistic jets from the source. It shows a peculiar shape characterized by a
significant excess near the orbital phase $\phi_{orb}= 0.25$, which is not seen in the softer 2-10 keV energy band. Such a shape is likely to be due to a complex
asymmetric structure of the funnel in a supercritical accretion disk in SS433. 
The orbital light curve at 40-60 keV demonstrates two almost equal 
bumps at phases $\sim 0.25$ and
$\sim 0.75$, most likely due to nutation effects of the accretion disk.   
The
change of the off-eclipse 18-60 keV X-ray flux with the precessional phase shows a double-wave form with strong primary maximum at $\psi_{pr}= 0$ and weak but significant
secondary maximum at $\psi_{pr}= 0.6$. A weak variability of the 18-60 keV flux in the middle of the orbital eclipse correlated with the disk precessional phase is also
observed. 
The joint analysis of the broadband (18-60 keV)
orbital and precessional light curves obtained by INTEGRAL 
confirms the presence of a hot extended corona in the central parts of the
supercritical accretion disk and constrain the 
binary mass ratio in SS433 in the range $0.5\gtrsim q\gtrsim 0.3$, 
confirming the black hole nature of the compact object. Orbital and precessional 
light curves in the hardest X-ray band 40-60 keV, which
is free from emission from thermal X-ray jets, are also best fitted by the same 
geometrical model with hot extended corona 
at $q\sim 0.3$, stressing the conclusions of the modeling
of the broad-band X-ray  orbital and precessional light curves. 
}

\FullConference{An INTEGRAL view of the high-energy sky (the first 10 years)
- 9th INTEGRAL Workshop and celebration of the 10th anniversary of the
- launch,\\
15-19 October 2012\\
Bibliotheque Nationale de France, Paris, France }

\begin{document}

\section{Introduction}
\label{sec:intro}

SS433 is a unique galactic 
steadily superaccreting microquasar with mildly relativistic ($v=0.26c$), 
precessing jets located at a distance of 5.5 kpc \cite{Margon84, Cher81, Cher88, CrHu81, Fab04}.
The system exhibits three photometric and
spectral periodicities related to precession
($P_\mathrm{prec}=162^d.5$), orbital ($P_\mathrm{orb}=13^d.082$) and
nutation ($P_\mathrm{nut}=6^d.28$) periods \cite{Goransk98}.
Despite of wealth of observations, the nature of the compact star in 
SS433 remains inconclusive. The presence
of absorption lines in the optical spectrum of the companion \cite{Gies02,hillwig08} suggests its spectral classification as $\sim$ A7Ib supergiant. 
Assuming these lines to be produced in the optical
star photosphere, their observed orbital Doppler shifts 
would correspond to the mass ratio of
compact ($M_x$) and optical ($M_v$) star $q=M_x/M_v\sim 0.3\pm
0.11$ and masses $M_x=(4.3\pm 0.8) M_\odot$, $M_v=(12.3\pm3.3) M_\odot$, respectively,
pointing to the black hole nature of the compact star. 

Modeling of all INTEGRAL eclipses of the source available before 2010 
\cite{Cher09} using a purely geometrical model 
yeilded independent constraints on   
the binary mass ratio $q=0.25-0.5$ with the most probable value $q=0.3$, suggesting
the mass of the compact companion $M_x\simeq 5.3 M_\odot$ and the optical star 
$M_v\simeq 17.7 M_\odot$ for the observed optical star mass function 
$f_v=0.268 M_\odot$. This places SS433 among black-hole high-mass X-ray binaries. 
Thus SS433 can be the only known example of galactic massive X-ray binary at an advanced
evolutionary stage \cite{Cher81, Cher88} with supercritical
accretion \cite{ShS73} onto a black hole \cite{Cher09}.
Its study in different spectral bands provides invaluable information for 
theory of evolution of binary star and the formation of relativistic jets.  

The basic picture of hard X-ray emitting regions, 
as emerged from analysis of X-ray data \cite{Fab04, Fil06, Cher03, Cher09, Krivosheev09}, includes 
hot X-ray jet propagating through a funnel in the supercritical 
accretion disk, filled with hot scattering medium (a corona).  
The X-ray spectrum of SS433 in the 3-100 keV range 
can be fitted by two-component model (thermal X-ray emission from 
the jet and thermal comptonization spectrum from corona) 
elaborated in \cite{Krivosheev09}.  The scattering corona 
parameters are: $T_{cor}\simeq 20$~keV, Thomson optical depth $\tau_T\simeq 0.2$
and mass outflow rate in the jet $\dot M_j=3\times 10^{19}$~g/s.
This parameters suggest the coronal electron number density around $5\times 10^{12}$ cm$^{-3}$, which is
typical in the wind outflowing with a velocity of $v\sim 3000$ km/s from a supercritical 
accretion disk with mass accretion rate onto the compact star $\dot M\sim 10^{-4}$
M$_\odot$/yr at distances $\sim 10^{12}$ cm from 
the center, where a Compton-thick photosphere is formed \cite{Fab04}.
The size of the disk photosphere was independently estimated
from measurements of fast optical aperiodic variability \cite{Burenin2010}. 

Here we analyze hard X-ray eclipses of SS433 near the T3 moment in 
combination with the precessional variability
as observed by INTEGRAL, and interpret them in terms of our multicomponent 
geometrical model (see \cite{Cher09} for more detail). 

\section{Composite X-ray light curve and jet nutation effect}

Previous dedicated INTEGRAL observations of hard X-ray eclipse are summarized in
Table \ref{t:obs}. They were all concentrated around precessional 
phase zero ('the T3 moment' in terms of the kinematic model of SS433 \cite{Margon84}), where the accretion disk is maximum opened to the observer
and the X-ray flux from the source is the highest. 

\begin{table*}
\caption{Dedicated observations of SS433 by INTEGRAL\label{t:obs}}
\footnotesize
\centering
\begin{tabular}{lllc}
\hline
Set&INTEGRAL orbits& Dates & Precessional phase $\psi_{pr}$\\
\hline
I&67-70 & May 2003 &0.001-0.06\\
II&555-556 & May 2007 &0.98-0.014\\
III&608-609 &October 2007 &0.956-0.99\\
IV&612-613 &October 2007 &0.030-0.064\\ 
V&722-723 &September 2008&0.057-0.091\\
VI&984     &November 2010&0.87-0.89\\ 
&987     &November 2010&0.93-0.94\\
VII& 1040-1041&April 2011&0.91-0.95\\
\hline 
\end{tabular}
\end{table*}

\begin{figure*}
\includegraphics[width=0.32\textwidth]{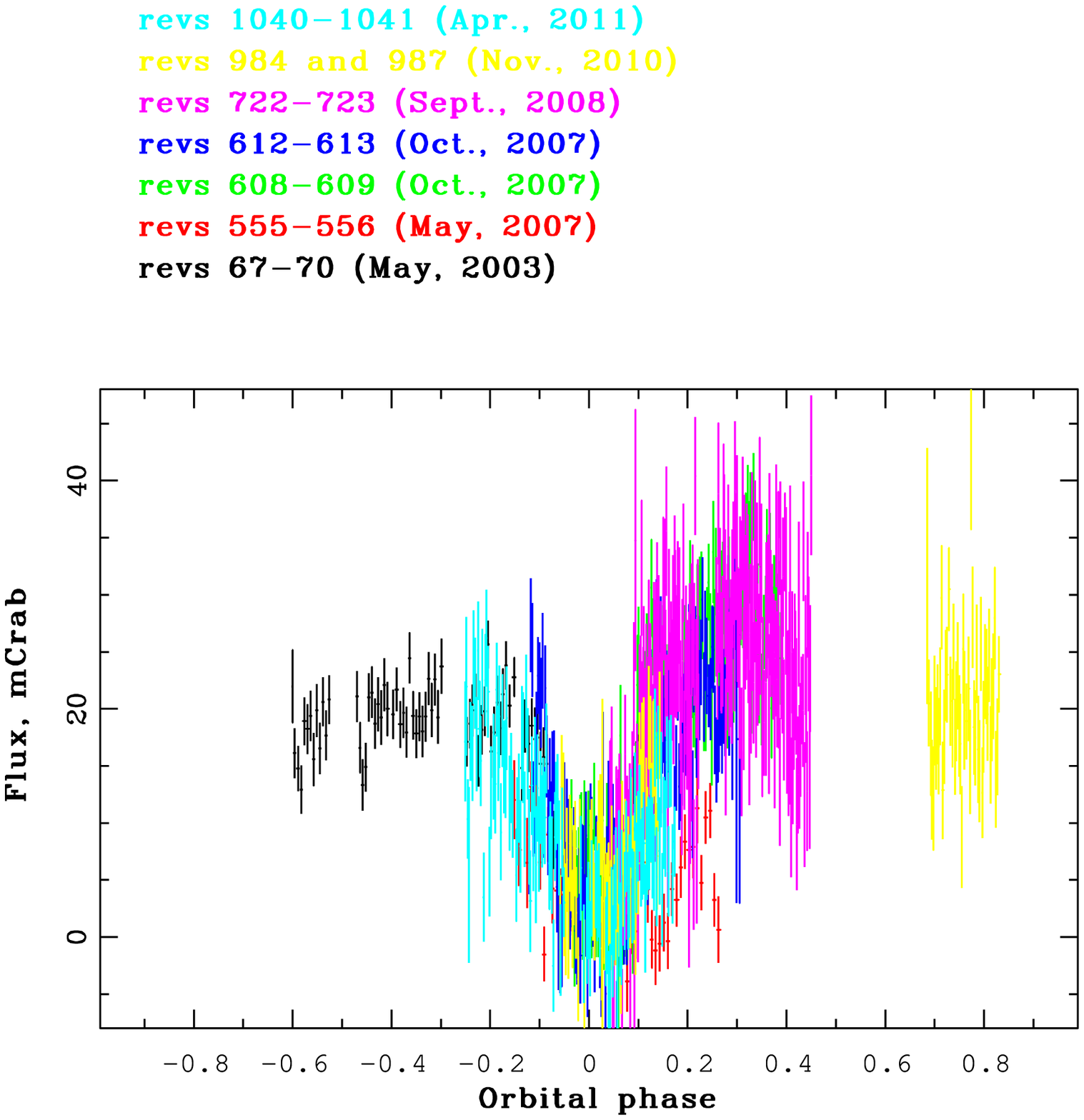}
\hfill
\includegraphics[width=0.32\textwidth]{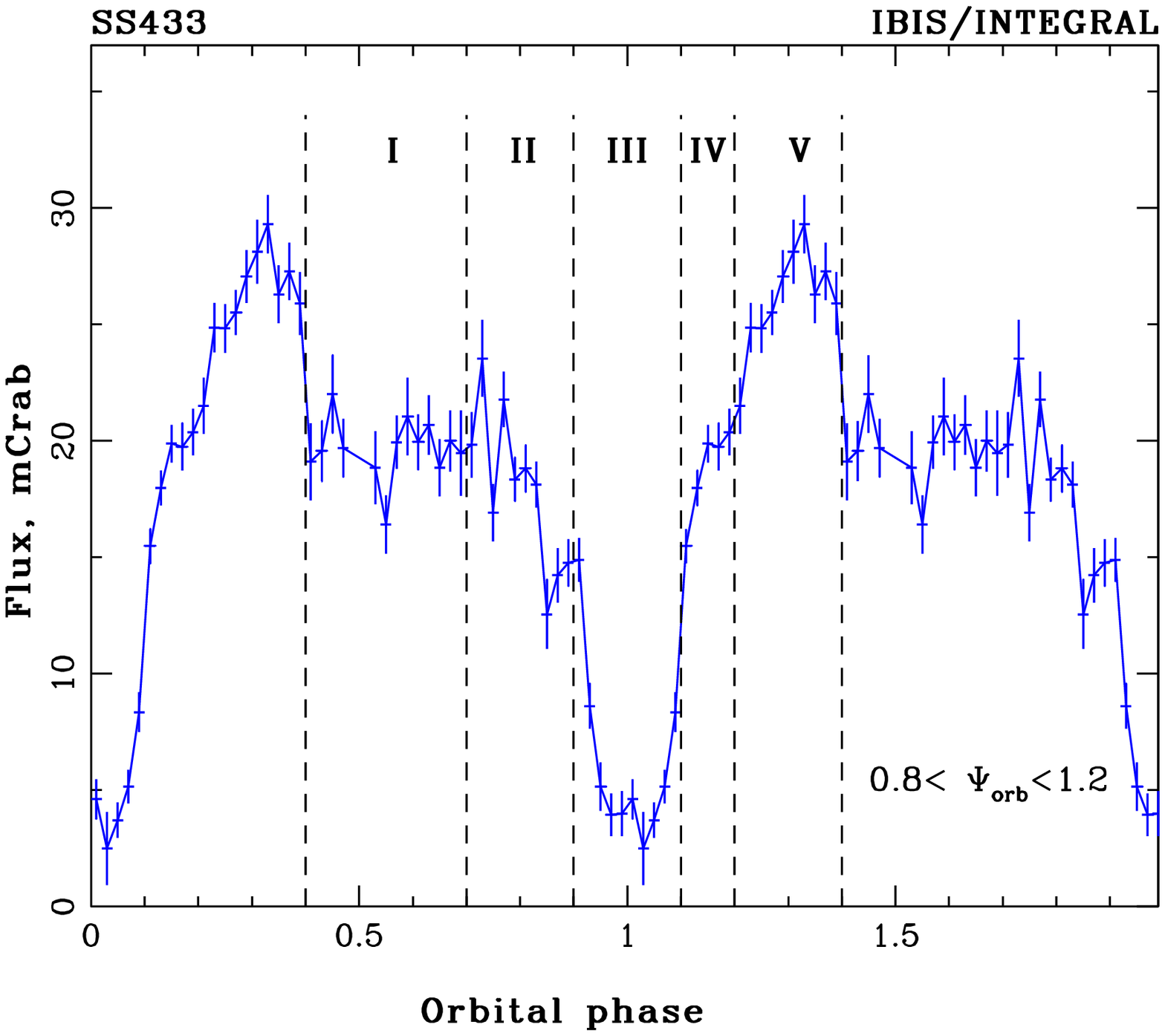}
\hfill 
\includegraphics[width=0.32\textwidth]{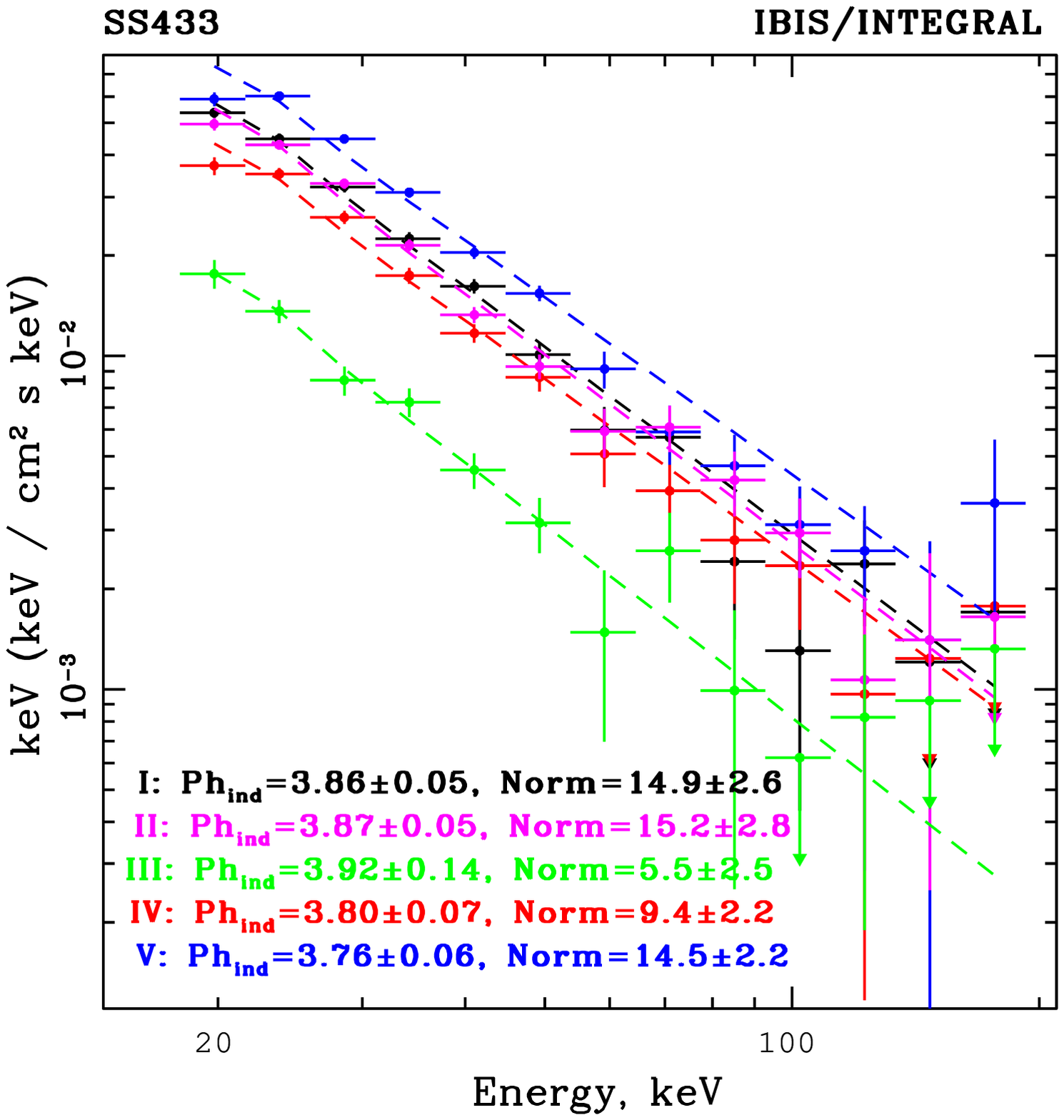}
\caption{From left to right: 1) Composite IBIS/ISGRI 18-60 keV X-ray eclipse 
light curve around zero precessional phase $\psi=0$. 2)  
Binned X-ray light curve ($\Delta \phi=0.02$) with  
phase intervals for spectral analysis. 3) Phase-resolved IBIS/ISGRI 
spectra of SS433 within chosen orbital phase intervals. 1-$\sigma$ errors
are shown.}
\label{f:lc}
\end{figure*}

\begin{figure*}

\includegraphics[width=0.45\textwidth]{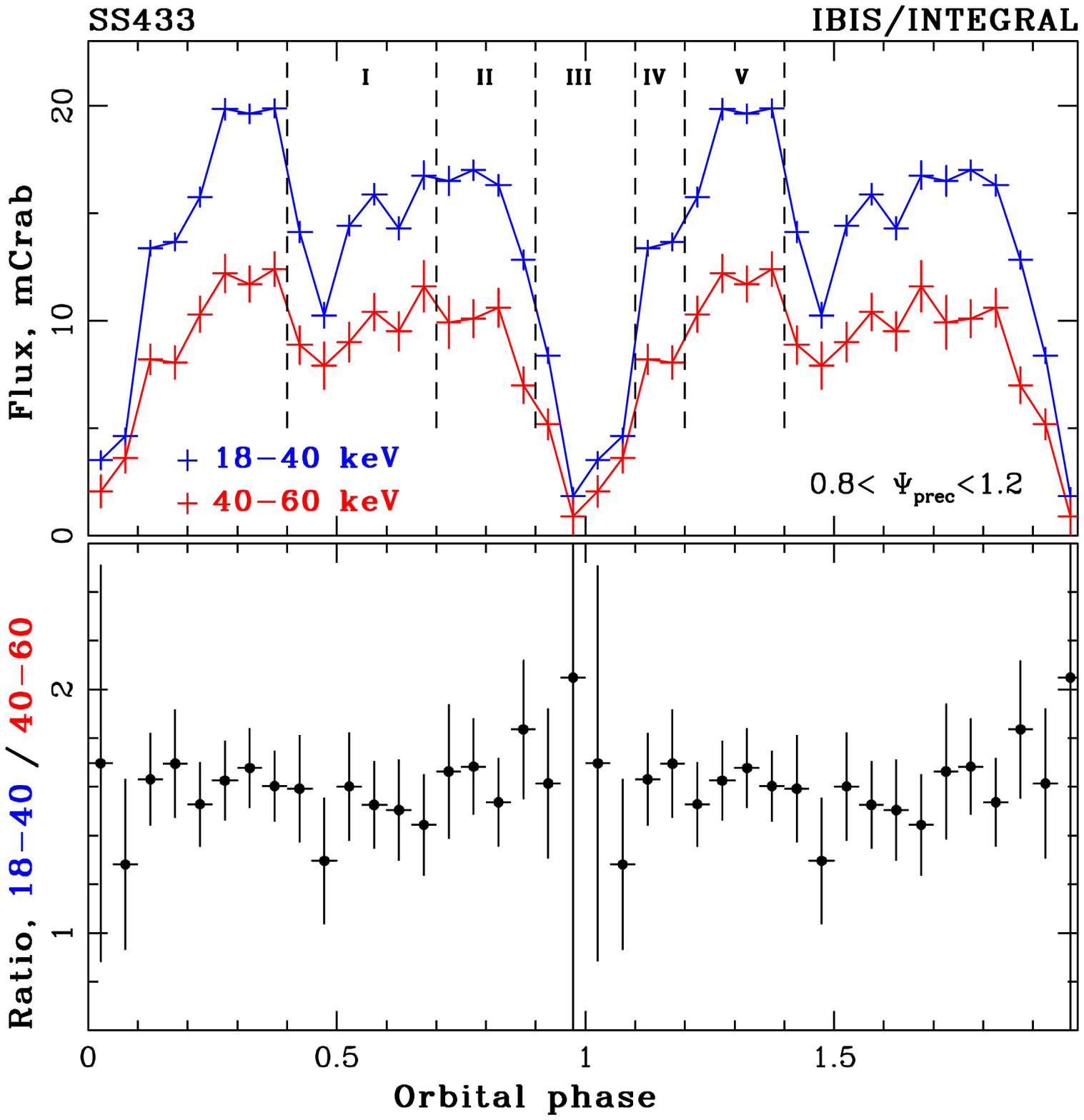}
\hfill
\includegraphics[width=0.45\textwidth]{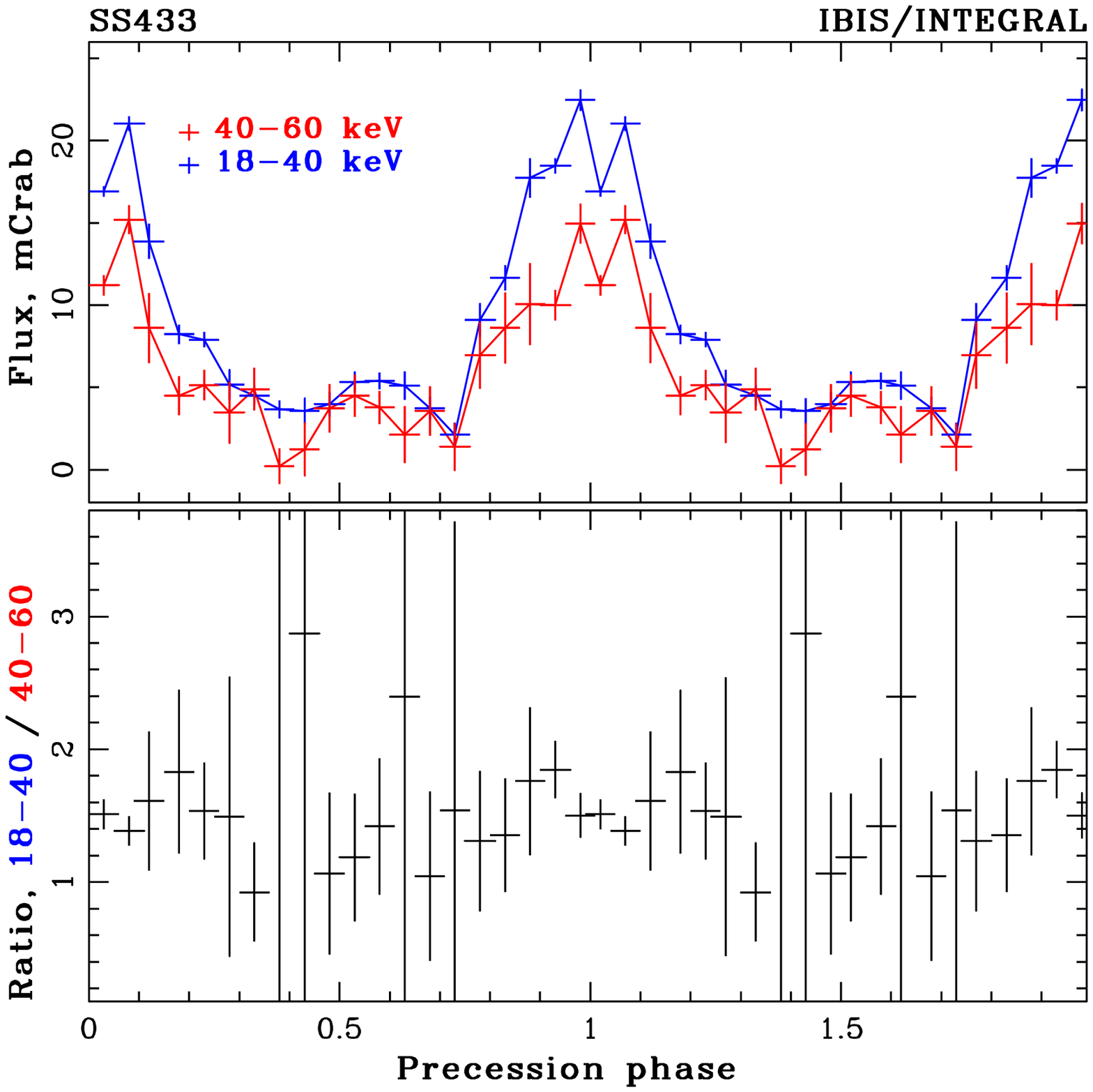}
\parbox[t]{0.47\textwidth}{\caption{IBIS/ISGRI 18-40 and 40-60 keV orbital light curves with hardness
ratio (bottom plot).}\label{f:horb}}
\hfill
\parbox[t]{0.47\textwidth}
{\caption{Precessional 18-40/40-60 keV light curves and hardness ratio.}\label{f:hprec}}
\end{figure*}

The composite IBIS/ISGRI 18-60 keV light curve of the primary X-ray eclipse
at precessional phase zero is shown in Fig. \ref{f:lc} (panel 1). Data were analyzed 
using the IKI INTEGRAL data processing code described in \cite{Mol04}. The
eclipse light curve averaged within orbital phase bins $\Delta \phi=0.02$ is 
shown in Fig. \ref{f:lc} (panel 2). 3$\sigma$-flux errors are indicated.
The 18-40 and 40-60 keV orbital and precessional light curves with the corresponding hardness ratios are shown in Fig. \ref{f:horb} and Fig. \ref{f:hprec}, respectively.  

A very significant flux excess at the orbital phase $\phi \sim 0.25$ is observed 
in the composite 18-60 light curve (Fig.  \ref{f:lc}, panel 2; Fig. \ref{f:horb}, upper light curve)
after the eclipse relative to the phase 0.75 before the eclipse. 
However, on the 40-60 keV light curve two maxima with similar amplitude 
are clearly seen at both orbital phases 0.25 and 0.75. These $\sim 10\%$ 
sine-like variability at twice the orbital period superimposed on the 
orbital light curve is mostly likely to be due to the jet nutation.
The tidal nutation occurs twice the
synodic period ($6.28$~d in the case of SS433). The effect 
must be maximal at the zero precessional phase when the 
binary system is observed in quadratures (i.e. at the binary phases 0.25 and 0.75). 
Due to nutation the jet changes its inclination
to the line of sight by 6 degrees, suggesting the change in the projection
area of the jet funnel $\Delta S/S\sim \Delta i_j/i_j=0.1$ (here $i_j\approx 
59^o$ is the jet inclinatin angle at the T3 precessional phase).    
Thus, the composite INTEGRAL 
light curve allowed us to see for the first time the jet nutation in the 40-60 keV band.

\noindent \textbf{2. Orbital phase-resolved spectroscopy}. 
The accumulated data allowed us to make for the first time 
the orbital phase-resolved X-ray spectroscopy. Five orbital phase intervals chosen for spectral analysis are shown by the vertical dashed lines in Fig. \ref{f:lc} (middle panel). The obtained X-ray spectra are shown in Fig. \ref{f:lc} (right panel). It is seen
that within errors they have an identical power-law shape with photon spectral index 
$\Gamma_{ph}\simeq 3.8$. Nevertheless, a tendency of the spectrum to get
harder at $\phi\sim 0.25$ (phase interval V in Fig. \ref{f:lc}, middle panel) than 
at $\phi_{orb}\sim 0.75$ (phase interval I in Fig. \ref{f:lc}, middle panel) is clearly 
seen. Note also that at the middle eclipse (phase interval III in Fig. \ref{f:lc}, middle panel) the spectrum gets softer. This is exactly what is expected in the jet 
nutation picture: during the disk-jet nutation the angle between the line of sight
and the jet axis changes by $\sim 6$ degrees, so the observer looks into 
the funnel less (at $\phi_{orb}=0.5$) or more  (at $\phi_{orb}=0.75$) deeper, thus 
observing cooler or hotter parts of the jet base.

\noindent \textbf{3. Precessional variability}. 
At a distance of 5.5 kpc to SS433, reliably derived from the kinematic
properties of moving emission lines \cite{Fab04}, the observed X-ray flux 18-60 keV 
corresponds to a maximum uneclipsed hard X-ray luminosity of 
$3\times 10^{35}$~erg/s. The precessional change of the X-ray flux is shown in 
Fig. \ref{f:hprec}.  To plot this Figure, all available observations of
SS433 by INTEGRAL with a total exposure time of about 8 Ms were used.  
This confirms the presence of 
a fairly broad region emitting in hard X-rays with size 
comparable to that of the accretion disk ($\sim 10^{12}$ cm), 
since (excluding orbital eclipses) 
the observed flux varies due to precession of the disk.
The both precessional light curves
show precessional variability and were used to constrain parameters of 
the hot corona (\cite{Cher09} and below). 

\section{Results}

\begin{figure*}
\includegraphics[width=0.24\textwidth]{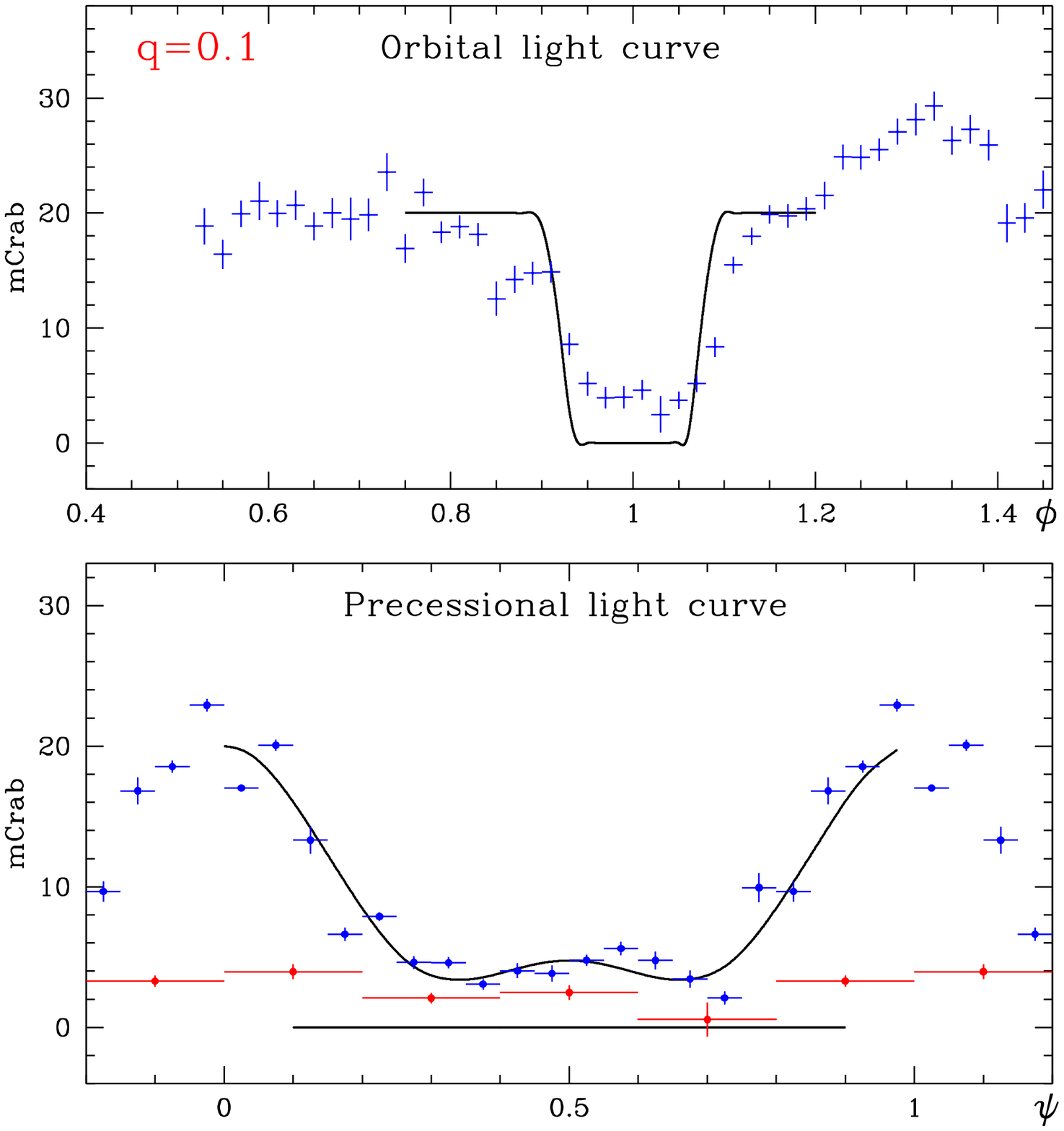} 
\hfill
\includegraphics[width=0.24\textwidth]{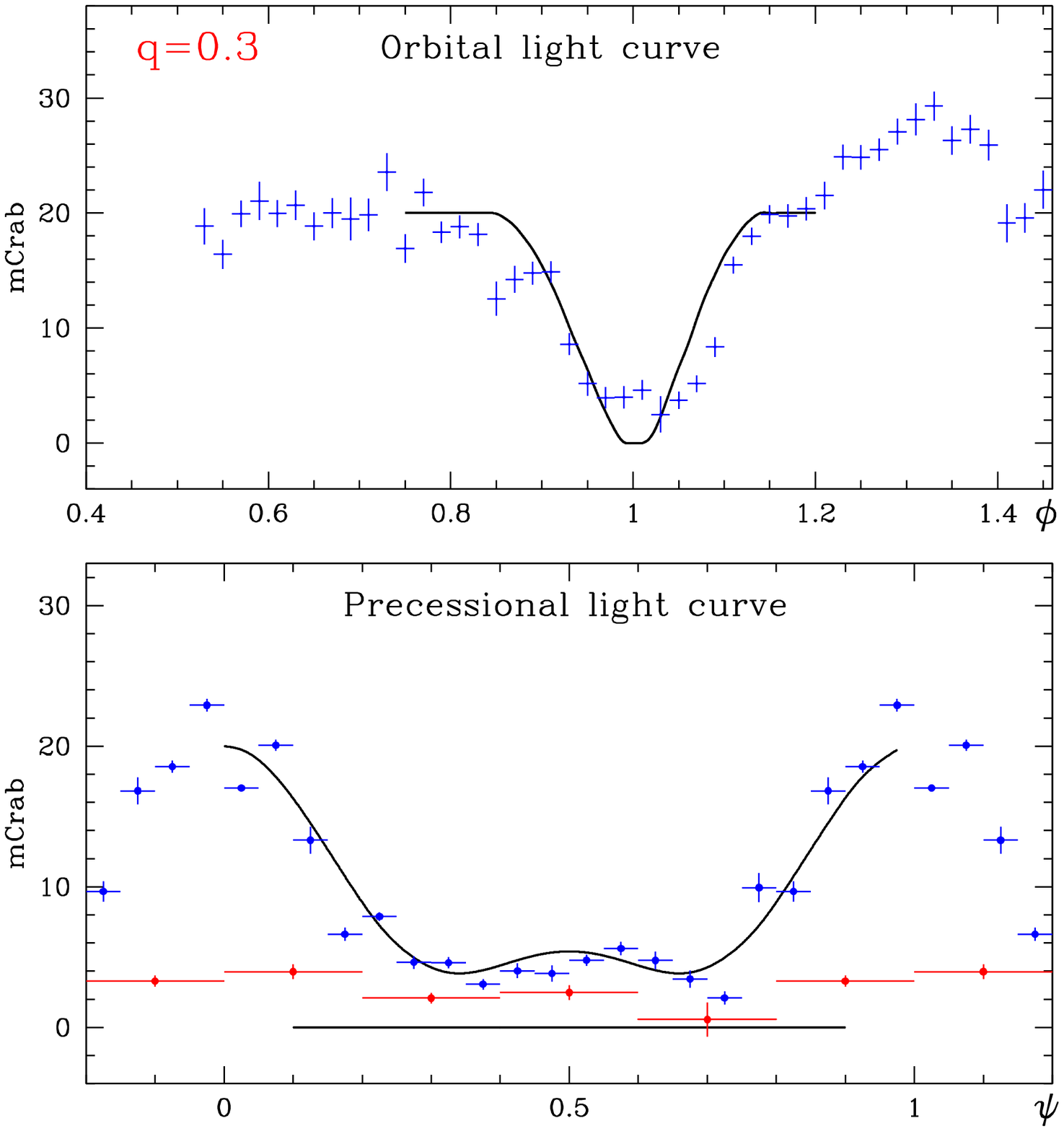} 
\hfill
\includegraphics[width=0.24\textwidth]{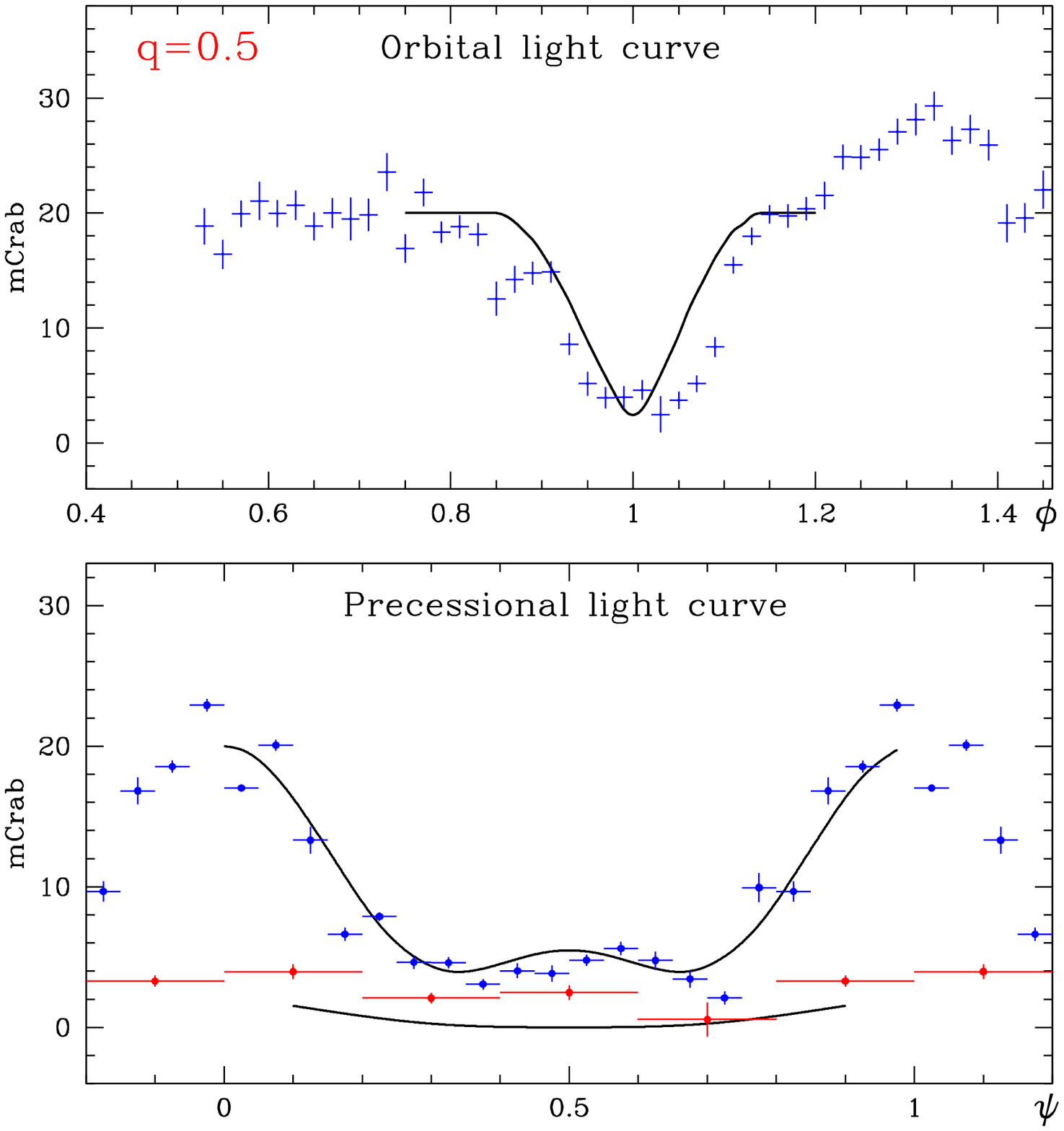}
\hfill
\includegraphics[width=0.24\textwidth]{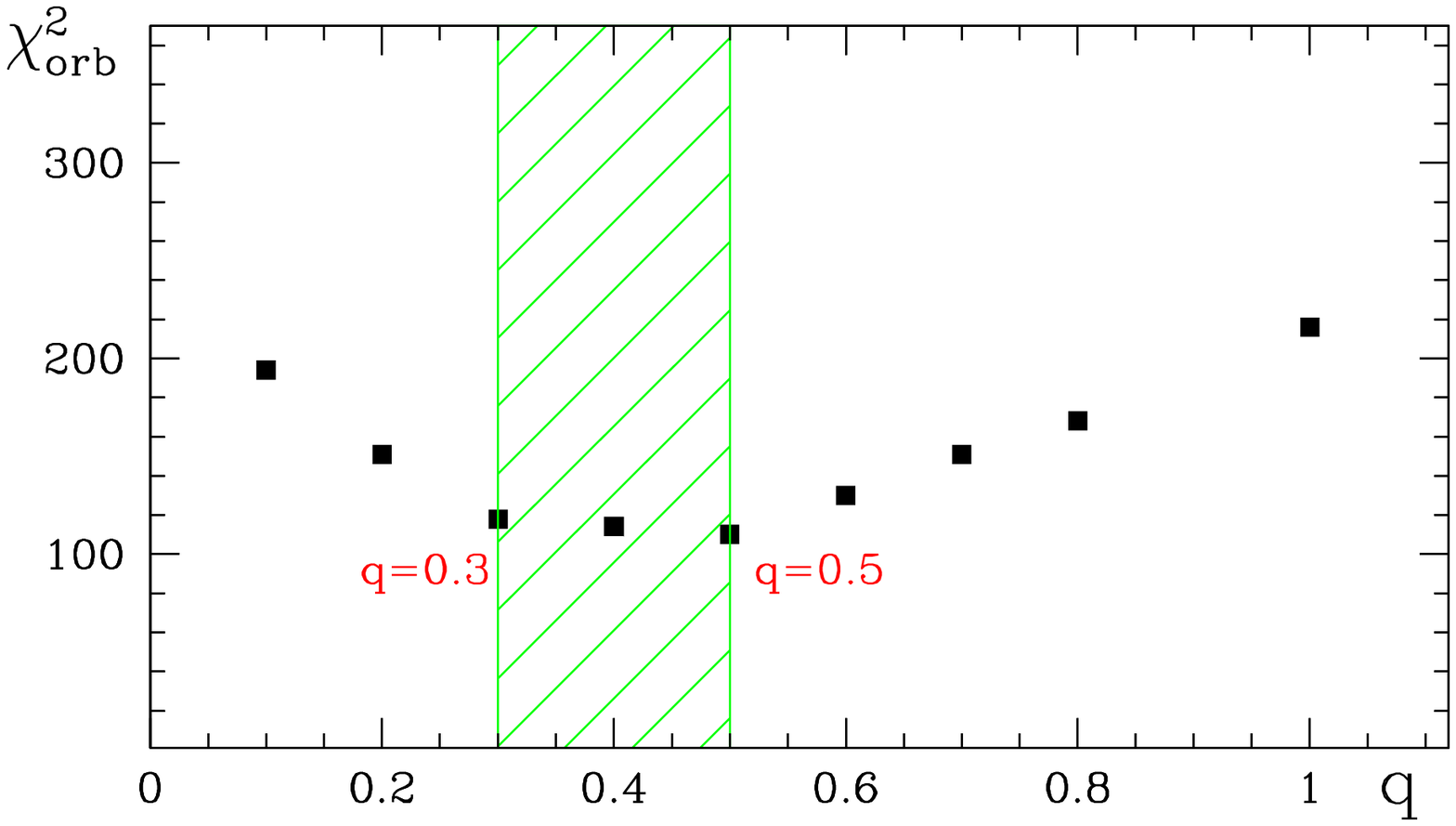}
\caption{Joint analysis of orbital (upper plots) and precessional (lower plots) 
18-60 keV light curves of SS433. On the bottom panels, precessional variability in the middle of eclipse ($0.95<\phi_{orb}<1.05$) are shown by red crosses. 
1). $q=0.1$, 'short jet' corona
($a_j=0.25$, $b_j=0.1$, 
$\omega=40^o$). 2) $q=0.3$, 'short jet' corona
($a_j=0.35$, $b_j=0.13$, 
$\omega=80^o$). 
3) $q=0.3$, 'short jet' corona
($a_j=0.35$, $b_j=0.13$, 
$\omega=80^o$). 4) $\chi^2$ for the orbital light curve for different mass ratios.}
\label{f:lcanalysis1860}
\end{figure*}

The INTEGRAL observations of SS433 provide three different light curves, which can be used to constrain the parameters of the system: 1) the orbital light curve, 2) the precessional light
curve out of eclipses, and 3) precessional light curve in the middle of the eclipses (see 
Fig. \ref{f:lcanalysis1860}, upper panels, and lower panels, 
blue and red crosses, respectively). 
We use a geometrical model described in detail in  \cite{Cher09}. Briefly, 
the model includes an optical star with mass $M_v$ filling its Roche lobe, 
a compact star with mass $M_x$ surrounded by an optically thick 
accretion disk with radius $a_d$, 
and a hot corona
which is modeled as a broad 'jet' parametrized by the part of ellipse 
with semi-axes $a_j$ and $b_j$ normalized to the binary orbital separation $a$. The elliptical corona 
is restricted by the cone with half-angle $\omega$.    
The shape and amplitude
of the precessional light curve constrain the height of the hot X-ray corona, while  
the orbital light curve restricts the accretion disk radius. 
At a given binary mass ratio, 
after finding the best parameters for the precessional variability, we 
can calculate the deviations of the model orbital light curve from the observed one.  
The results of the joint analysis of the orbital and precessional 18-60 keV 
light curves of SS433 are shown in Fig. \ref{f:lcanalysis1860}.

\begin{figure*}
\includegraphics[width=0.3\textwidth]{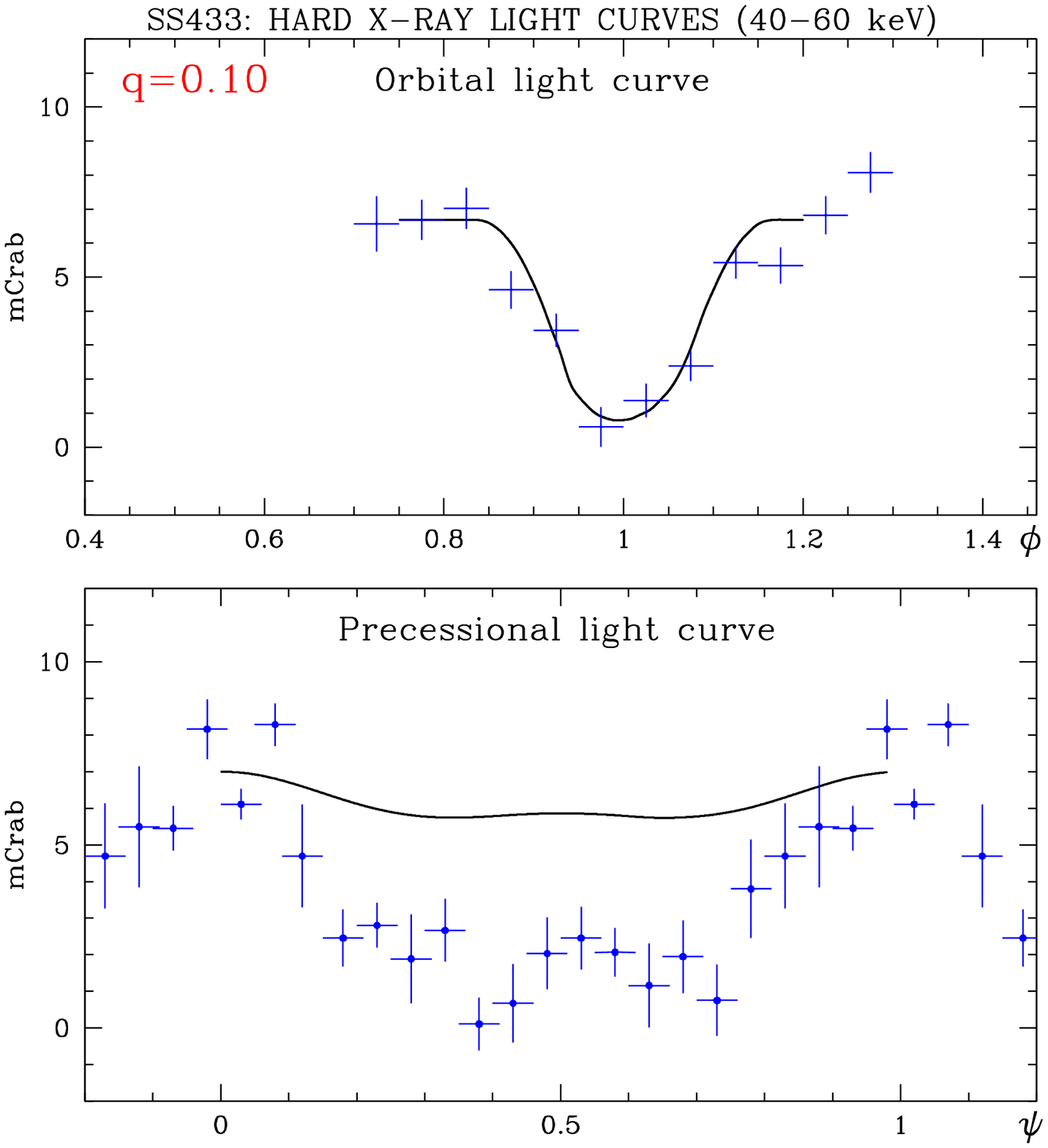}
\hfill
\includegraphics[width=0.3\textwidth]{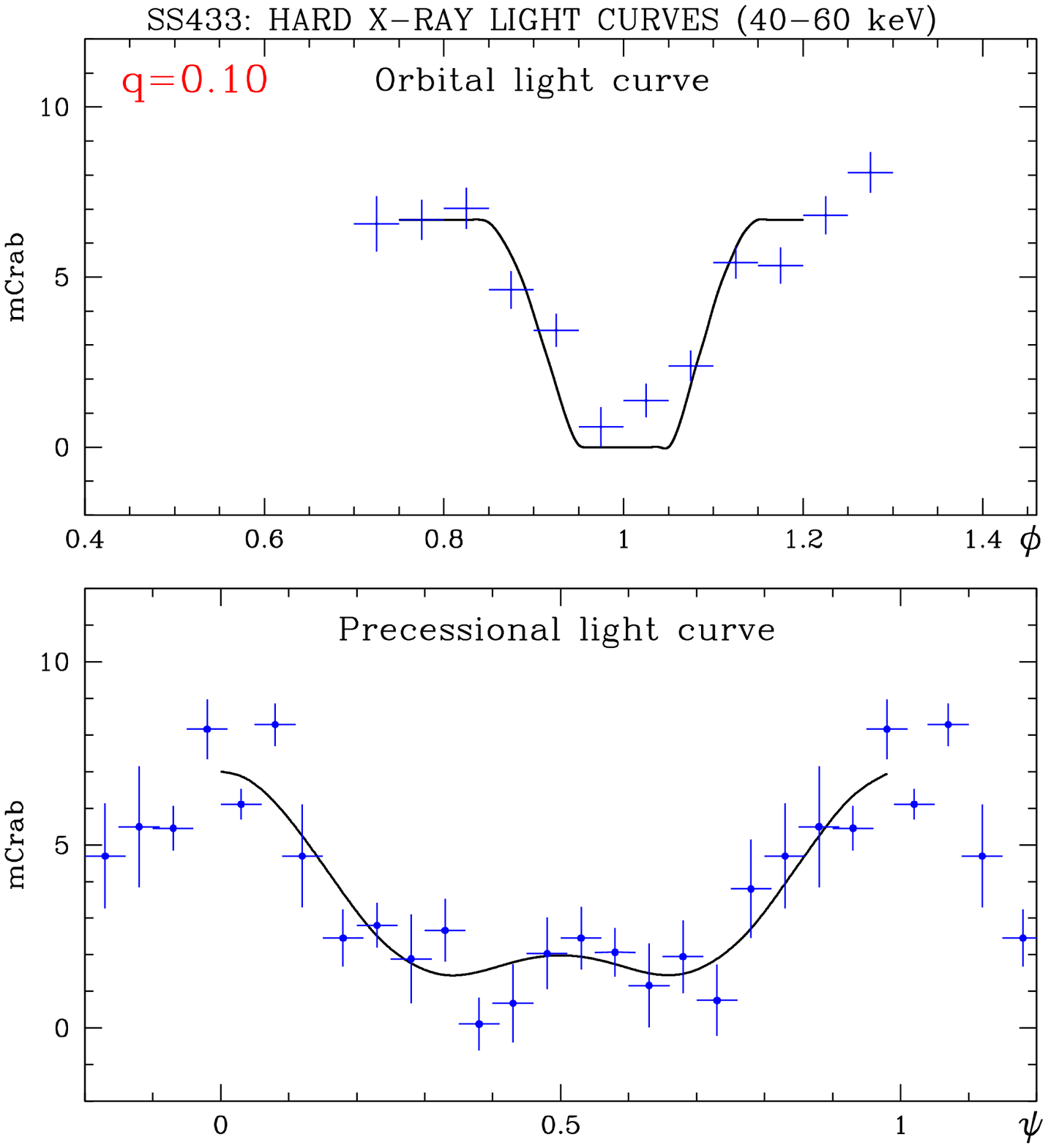} 
\hfill
\includegraphics[width=0.3\textwidth]{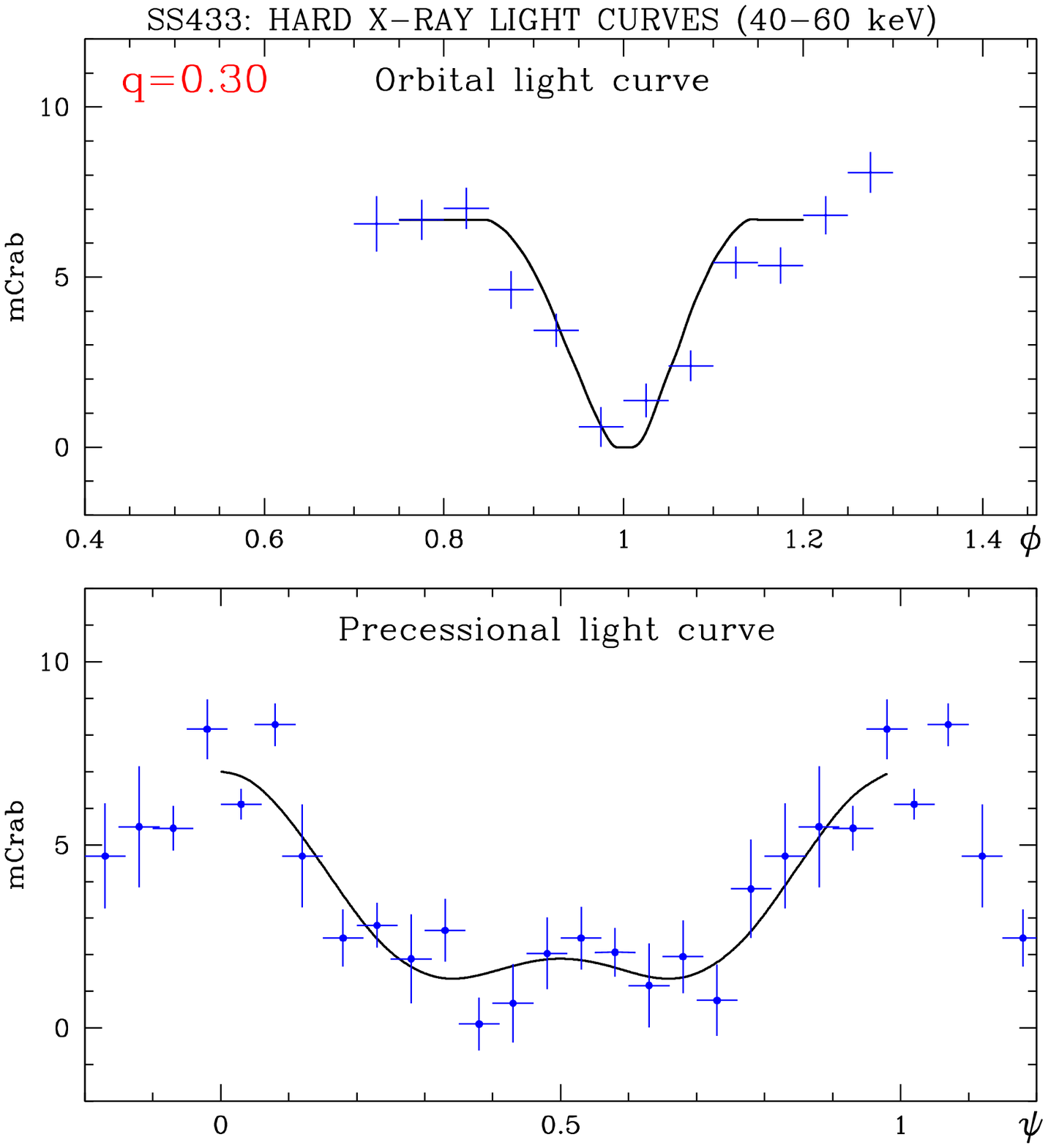} 
\caption{Joint analysis of orbital (upper plots) and precessional (lower plots) 
40-60 keV light curves of SS433. Left panel: $q=0.1$, 'long jet' corona ($a_j=0.25$, $b_j=0.55$, 
$\omega=80^o$); only orbital 
light curve can be reproduced. Middle panel: $q=0.1$, 'short jet' corona
($a_j=0.25$, $b_j=0.1$, 
$\omega=80^o$). Both 
orbital and precessional light curvs can be fitted, but the total eclipse (plateau at zero flux) appears (unobserved). Right panel: $q=3$, 'short jet' corona
($a_j=0.35$, $b_j=0.13$, 
$\omega=80^o$); both orbital and precessional light curves are well reproduced.}
\label{f:lcanalysis}
\end{figure*}

To avoid contamination from thermal X-ray emission from relativistic jets, 
we repeated the analysis using only hard X-ray 40-60 keV light curves 
shown in Fig. \ref{f:horb} and Fig. \ref{f:hprec}. 
The result for different mass ratios $q=M_x/M_v$ is presented in  
Fig. \ref{f:lcanalysis}. It is seen that the hard
X-ray orbital and precessional light curves can be simultaneously reproduced 
by the geometrical model for the binary mass ratio $q\gtrsim 0.3$ (at smaller mass ratios a
plateau corresponding to the total eclipse of the hot corona by the optical star appears 
in the orbital light curve, which is not observed), in agreement with the analysis of
the broadband 18-60 keV orbital and precessional light curves shown in Fig. \ref{f:lcanalysis1860}.

\section{Conclusions}

1) INTEGRAL observations of SS433 in hard X-rays 18-60 keV allowed us for the first time 
to make orbital-resolved spectroscopy of the X-ray eclipse in the precessional phase
corresponding to the maximum opening angle of the disk. The hard X-ray continuum is fitted with a
power-law with photon index $\Gamma\approx 3.8$ which does not 
significantly change across the eclipse, 
suggesting the origin of this emission as being due to scattering in hot corona 
surrounding the funel around the jets in a supercritical accretion disk in SS433.

2) For the first time, joint analysis of hard X-ray (40-60 keV) orbital and precessional light curves has been performed. This analysis independently 
confirms our previous result \cite{Cher09} that 
the low value of the mass ratio $q=M_x/M_v$ in SS433 cannot reproduce the observed orbital
and precessional light curves. With the existing estimates of the 
mass function of the compact star, the most likely value $q\sim 0.3$ 
points to the black hole nature of the compact star in SS433. 

3) The shape of the hard X-ray orbital light curve 40-60 keV demonstrates two humps at
around orbital phases 0.25 and 0.75, likely due to the nutation effects in 
SS433. 

New INTEGRAL observations of SS433 at different precessional phases will
be used to further constrain physical parameters of this unique galactic microquasar.

\end{document}